\documentclass{article}
\usepackage{spconf,amsmath,graphicx}
\usepackage{url}
\usepackage{graphicx}
\usepackage{amsmath}   
\usepackage{amsfonts}  
\usepackage{bm}        
\usepackage{booktabs}  
\usepackage{multirow}  

\title{What's All the FUSS About Free Universal Sound Separation Data?}
\name{
\begin{tabular}{@{}c@{}}
Scott Wisdom$^1$, Hakan Erdogan$^1$, Daniel P. W. Ellis$^1$, Romain Serizel$^2$, Nicolas Turpault$^2$,\\
Eduardo Fonseca$^3$, Justin Salamon$^4$, Prem Seetharaman$^5$, John R. Hershey$^1$
\end{tabular}
}
\address{
$^1$Google Research, United States\\
$^2$Universit{\'e} de Lorraine, CNRS, Inria, Loria, Nancy, France \\
$^3$Music Technology Group, Universitat Pompeu Fabra, Barcelona, Spain \\
$^4$Adobe Research, San Francisco, United States \\
$^5$Descript, Inc., San Francisco, United States
}
\begin{document}
\ninept
\maketitle
\begin{abstract}
We introduce the \emph{Free Universal Sound Separation} (FUSS) dataset, a new corpus for experiments in separating mixtures of an unknown number of sounds from an open domain of sound types.  The dataset consists of 23 hours of single-source audio data drawn from 357 classes, which are used to create mixtures of one to four sources.
To simulate reverberation, an acoustic room simulator is used to generate impulse responses of box shaped rooms with frequency-dependent reflective walls.
Additional open-source data augmentation tools are also provided to produce new mixtures with different combinations of sources and room simulations.  
Finally, we introduce an open-source baseline separation model, based on an improved time-domain convolutional network (TDCN++), that can separate a variable number of sources in a mixture. This model achieves 9.8 dB of scale-invariant signal-to-noise ratio improvement (SI-SNRi) on mixtures with two to four sources, while reconstructing single-source inputs with 35.5 dB absolute SI-SNR.  We hope this dataset will lower the barrier to new research and allow for fast iteration and application of novel techniques from other machine learning domains to the sound separation challenge.
\end{abstract}
\begin{keywords}
Universal sound separation, variable source separation, open-source datasets, deep learning
\end{keywords}
\section{Introduction}
Hearing is often confounded by the problem of interference: multiple sounds can overlap and obscure each other, making it difficult to selectively attend to each sound.  In recent years this problem has been addressed by using deep neural networks to extract sounds of interest, separating them from a mixture.   Sound separation has made significant progress by focusing on constrained tasks, such as separating speech versus non-speech, or separating one speaker from another in a mixture of speakers, often with assumed prior knowledge of the number of sources to be separated.   However human hearing seems to require no such limitations, and recent work has engaged in \emph{universal sound separation}: the task of separating mixtures into their component sounds, regardless of the number and types of sounds.

One major hurdle to training models in this domain is that even if high-quality recordings of sound mixtures are available, they cannot be easily annotated with ground truth. High-quality simulation is one approach to overcome this limitation. Achieving good results requires supervised training using ground-truth source signals, drawn from a diverse set of sounds, and mixed using a realistic room simulator.  Although previous efforts have created open-domain data \cite{kavalerov2019universal}, the number of sources was still considered known, and the impact on the field was limited by proprietary licensing requirements.

To make such data widely available, we introduce the \emph{Free Universal Sound Separation} (FUSS) dataset.
FUSS relies on CC0-licensed audio clips from \url{freesound.org}.  We developed our own room simulator that generates the impulse response of a box shaped room with frequency-dependent reflective properties given a sound source location and a microphone location. As part of the dataset release, we also provide pre-calculated room impulse responses used for each audio sample along with mixing code, so the research community can simulate novel audio without running the computationally expensive room simulator.

Finally, we have released a masking-based baseline separation model, based on an improved time-domain convolutional network (TDCN++), described in our recent publications \cite{kavalerov2019universal, tzinis2019improving}. On the FUSS test set, this model achieves 9.8 dB of scale-invariant signal-to-noise ratio improvement (SI-SNRi) on mixtures with two to four sources, while reconstructing single-source inputs with 35.5 dB SI-SNR.

Source audio, reverb impulse responses, reverberated mixtures and sources, 
and a baseline model checkpoint are available for download\footnote{\url{https://zenodo.org/record/4012661}}. Code for reverberating and mixing audio data and for training the released model is available on GitHub\footnote{\url{https://git.io/JTusI}}.

The dataset was used in the DCASE 2020 challenge, as a component of the Sound Event Detection and Separation task. The released model served as a baseline for this competition, and a benchmark to demonstrate progress against in future experiments.

\begin{figure*}[t]
    \centering
    \includegraphics[width=0.49\linewidth]{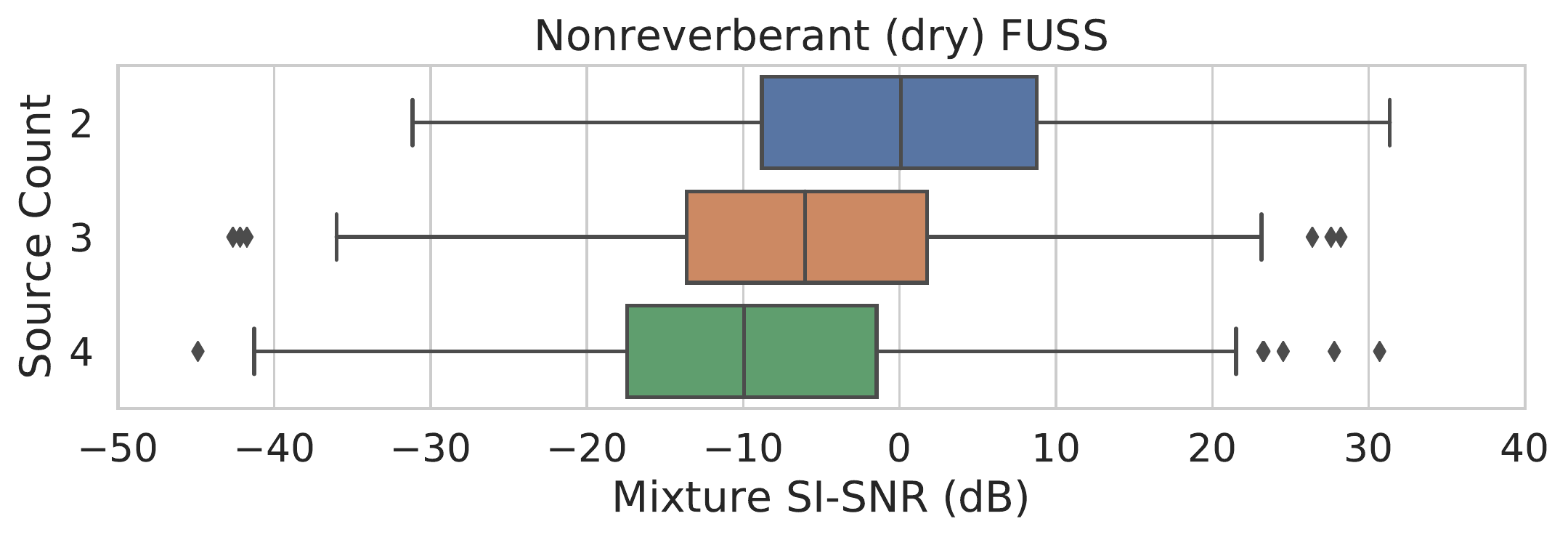}
    \includegraphics[width=0.49\linewidth]{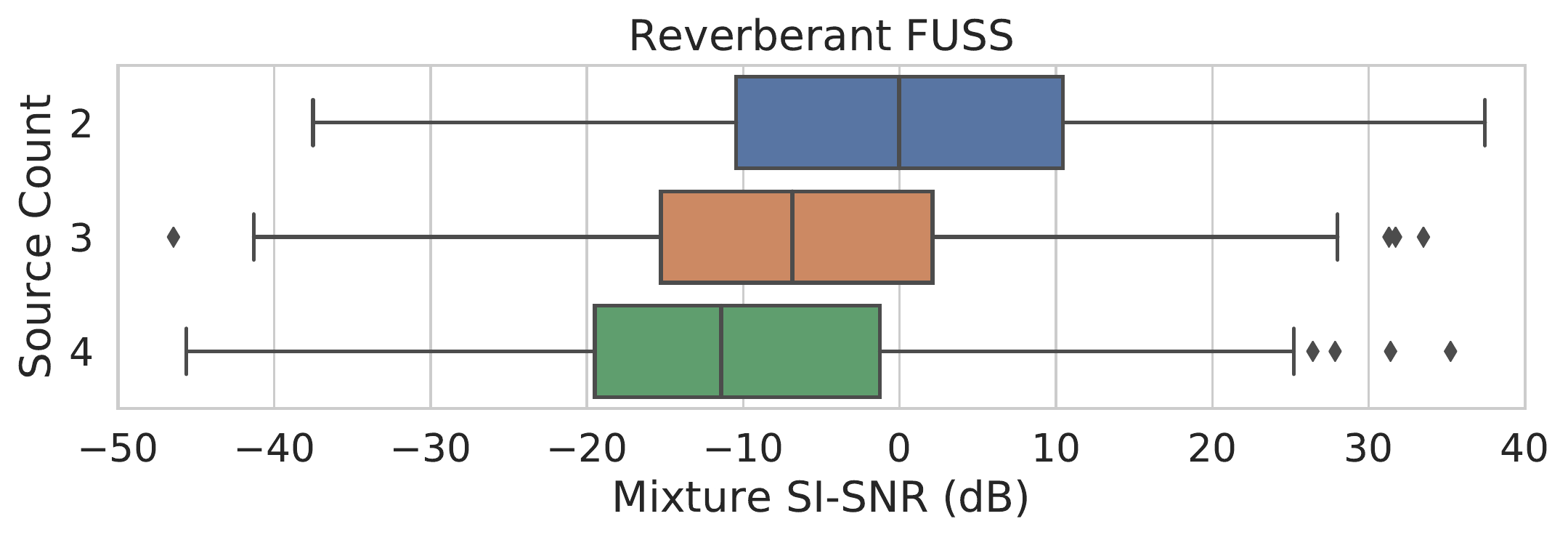}
    \vspace{-10pt}
    \label{fig:boxplot}
    \caption{Box plots for input SI-SNR for each source count on dry (left) and reverberant (right) FUSS.}
    \vspace{-10pt}
\end{figure*}

We hope this dataset will lower the barrier to new research, and particularly will allow for fast iteration and application of novel techniques from other machine learning domains to the sound separation challenge.
This paper provides three main contributions:
\begin{enumerate}
    \item We describe a new free and open-source dataset for universal sound separation, which is the largest to date in terms of both amount of data (23 hours of single-source audio data) and number of classes (357), and includes a variety of conditions, including variable source number (1-4) and reverberation.
    \item We propose new loss functions and evaluation metrics for models that can separate variable numbers of sources.
    \item We provide a baseline implementation of an open-domain separation system designed for variable numbers of sources, to serve as a benchmark for future work.  
\end{enumerate}

\section{Relation to prior work}
Only recently was open-domain (i.e. hundreds of sound classes) universal sound separation shown to be feasible \cite{kavalerov2019universal}, though only with a fixed number of sources, and on proprietary data with restrictive licensing. Zadeh et al.~\cite{zadeh2019wildmix} constructed a small (less than $1$ hour) dataset with 25 sound classes and proposed a transformer-based model to separate a fixed number of sources. Tzinis et al.~\cite{tzinis2020two} performed separation experiments with a fixed number of sources on the 50-class ESC-50 dataset \cite{piczak2015esc}. Other papers have leveraged information about sound class, either as conditioning information or as  as a weak supervised signal \cite{pishdadian2019finding, tzinis2019improving, kong2020source}.

In terms of variable source separation, Kinoshita et al.~\cite{kinoshita2018listening} proposed another approach for handling variable numbers of sources, where the separation network is made recurrent over sources. One drawback of this approach compared to ours is that the source-recurrent network needs to be run $N$ times to separate $N$ sources, while our proposed network only need to run once.   More recently, Luo and Mesgarani~\cite{luo2020separating} proposed a separation model trained to predict the mixture for inactive output sources. In contrast, our approach trains the separation model to predict zero for inactive sources, making it easier to determine when sources are inactive.

One typical application of universal sound separation is sound event detection (SED). In real scenarios SED systems commonly have to deal with complex soundscapes with possibly overlapping target sound events and non-target interfering sound events. Approaches have been developed to deal with overlapping target sound events using multilabeled (foreground vs background) training sets~\cite{salamon2015feature}, sets of binary classifiers~\cite{mesaros2016tut}, factorization techniques on the input of the classifier~\cite{benetos2016detection,bisot2017overlapping}, or exploiting spatial information when available~\cite{adavanne2018multichannel}. However, none of these approaches explicitly solved the problem of non-target events. Sound separation can be used for SED by first separating the component sounds in a mixed signal and then applying SED on each of the separated tracks~\cite{heittola2013supervised,kong2020source,serizel_2020,Huang2020,Cornell2020}.

\section{Data Preparation}

The audio data is sourced from a subset of FSD50K \cite{fonseca2020fsd50k}, a sound event dataset composed of \url{freesound.org} content annotated with labels from the AudioSet ontology \cite{AudioSet}. Audio clips are of variable length ranging from 0.3 to 30s, and labels were gathered through the Freesound Annotator \cite{fonseca2017freesound}.
Sound source files were selected which contain only one class label, so that they likely only contain a single type of sound. After also filtering for Creative Commons public domain (CC0) licenses, we obtained about 23 hours of audio, consisting of 12,377 source clips useful for mixing.  

To create mixtures, these sounds were split by uploader and further partitioned into 7,237 for training, 2,883 for validation, and 2,257 for evaluation.  Using this partition we created 20,000 training mixtures, 1,000 validation mixtures, and 1,000 evaluation mixtures, each 10s in length.   The sampling procedure is as follows.  For each mixture, the number of sounds is chosen uniformly at random in the range of one to four.   Every mixture contains zero to three \emph{foreground} source events, and one \emph{background} source, which is active for the entire duration.  The foreground and background sounds are sampled, using rejection sampling, such that each source in a mixture has a different sound class label.
For foreground sounds, a sound source file less than 10s long is sampled, and the entire sound is placed uniformly at random within the 10s clip.  For the background source, a file longer than 10s is sampled uniformly with replacement, and a 10s segment is chosen uniformly at random from within the file.  Choosing one longer source as a background for each mixture does introduce some bias in the class distribution of examples (examples of the biased classes include natural sounds such as wind, rain, and fire and man-made sounds such as piano, engine, bass guitar, and acoustic guitar), but it has the benefit of avoiding an unrealistic pattern of pure silence regions in every example.  Note that the background files were not screened for silence regions, and hence some of the background sounds may still contain significant periods of silence. Figure \ref{fig:boxplot} shows the distribution of input SI-SNR for examples with 2, 3, and 4 sources. Table \ref{tab:overlap} shows the proportion of local overlap, as measured with non-overlapping 25 ms windows. From this table it is clear that the background sources are active most of the time $81\%$), and mixtures with two or more sources contain sounds that do not always completely overlap.

\begin{table}[h]
    \vspace{-10pt}
    \centering
    \caption{Local overlap amount ($\%$) per source count.}
    \label{tab:overlap}
    \begin{tabular}{crrrrrrrrrr}
& \multicolumn{5}{c}{\bf Dry FUSS}
& \multicolumn{5}{c}{\bf Rev FUSS}
\\
\cmidrule(lr){2-6} \cmidrule(lr){7-11}
{\bf Count} 
& 0 & 1 & 2 & 3 & 4
& 0 & 1 & 2 & 3 & 4
\\
\midrule
1
& 19 & 81 & & &
& 23 & 77 & & &
\\
2
& 13 & 63 & 24 & &
& 20 & 59 & 21 & &
\\
3
& 8 & 47 & 36 & 9 &
& 11 & 45 & 35 & 8 &
\\
4
& 7 & 36 & 34 & 20 & 4
& 10 & 35 & 33 & 19 & 3
\\
\bottomrule
    \end{tabular}
\end{table}

The mixing procedure is implemented using Scaper \cite{salamon2017scaper}\footnote{We used Scaper 1.6.0 to generate FUSS.}. Scaper is a flexible tool for creating source separation datasets -- sound event parameters such as SNR, start time, duration, data augmentation (e.g., pitch shifting and time stretching), among others, can be sampled from user-defined distributions, allowing us to programmatically create and augment randomized, annotated soundscapes.   This allows FUSS to easily be extended by mixing larger amounts of training data and adding new source data and mixing conditions~\cite{Turpault2021}.  

\begin{figure*}[t]
    \centering
    \includegraphics[width=0.8\linewidth]{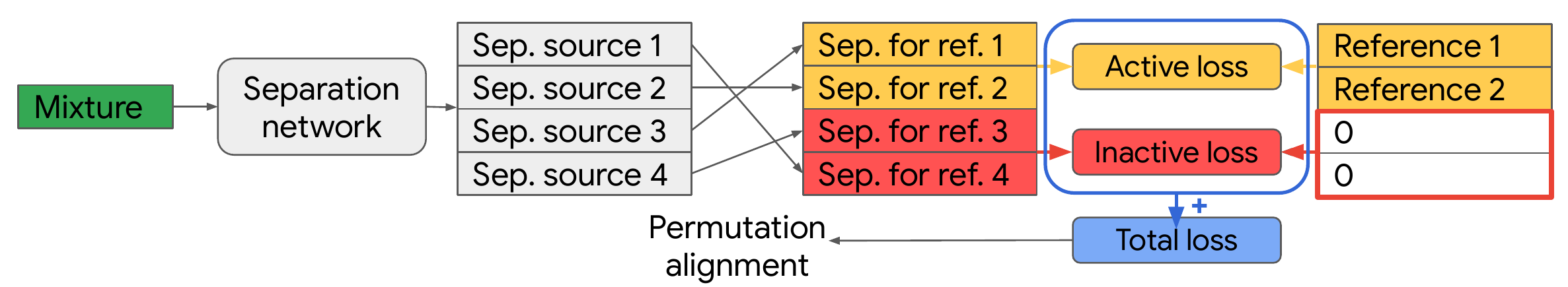}
    \vspace{-10pt}
    \caption{Variable source separation for a separation model with $M=4$  outputs and input mixture with $M_a=2$ active references. }
    \label{fig:vss}
    \vspace{-10pt}
\end{figure*}

The room simulation is based on the image method with frequency-dependent wall filters and is described in \cite{wang2020sequential}.
A simulated room with width between 3-7 meters, length between 4-8 meters, and height between 2.13-3.05 meters is sampled for each mixture, with a random microphone location, and the sources in the clip are each convolved with an impulse response from a different randomly sampled location within the simulated room. The locations are sampled uniformly, with a minimum distance of 20 cm between each source and all the microphones. During impulse response generation, all source image locations are jittered by up to 8 cm in each direction to avoid the sweeping echo problem \cite{de2015modeling}. The wall materials of each room is chosen randomly from common materials with frequency-dependent acoustic reflectivities, where we also randomly use a gain factor between 0.5 and 0.95 to increase variability of RIRs and better match real room impulse responses. We generate 20,000 train, 1000 validation, and 1000 test rooms.







\section{Baseline Model}

The baseline model uses a time-domain convolutional network (TDCN++) \cite{kavalerov2019universal, tzinis2019improving, wang2019alternating, wisdom2020unsupervised} that incorporates several improvements over the Conv-TasNet model \cite{luo2019conv}. These improvements include feature-wise layer normalization over frames instead of global normalization, longer-range skip-residual connections, and a learnable scaling parameter after each dense layer initialized to an exponentially
decaying scalar equal to $0.9^\ell$, where $\ell$ is the layer or block index. Using a short-time Fourier transform (STFT) with 32 ms window and 8 ms hop, input audio is transformed to a complex spectrogram. The magnitude of this spectrogram is fed to the TDCN++ network, which produces $M$ masks via a fully-connected output layer. These masks are multiplied with complex spectrogram input, and initial separated sources $\underline{\bf s}$ are produced by applying an inverse STFT. Finally, a mixture consistency layer \cite{wisdom2018consistency} is applied to these initial separated source waveforms:
\begin{equation}
    \hat{s}_{m}
    =
    \underline{s}_{m}
    +
    \frac{1}{M}
    (
        {x} - \sum_{m'} \underline{s}_{m'}
    ),
\label{eq:mixcon}
\end{equation}
which projects the separated sources such that they sum up to the original input mixture $x$. Since we know that FUSS mixtures contain up to 4 sources, we choose to use $M=4$.

\subsection{Variable Source Separation}

This baseline model is able to separate mixtures with a variable numbers of sources. To accomplish this, we propose a new loss function, as illustrated in Figure \ref{fig:vss}. Assume we have a training mixture ${x}$ with $M_a$ reference sources, where $M_a$ can be less than the number of output sources $M$. Thus, the training mixture has $M_a \leq M$ active reference sources ${\bf s}\in\mathbb{R}^{M_a \times T}$, while the separation model produces separated sources $\hat{\bf s}\in\mathbb{R}^{M\times T}$. For such an example, we use the following permutation-invariant training loss:
\begin{equation}
\begin{aligned}
    \mathcal{L}({\bf s}, \hat{\bf s})
    =
    \min_{  {\pi \in \Pi}}
    \Big[&
        \sum_{m_a=1}^{M_a}
        \mathcal{L}_\mathrm{SNR}(
            {s}_{m_a}, \hat{s}_{\pi(m_a)}
        )
    \\
    &+
    \sum_{m_0=M_a+1}^{M}
    \mathcal{L}_0(
        {x}, \hat{s}_{\pi(m_0)}
    )
    \Big],
    \end{aligned}
\end{equation}
where the active per-source loss is negative SNR with a threshold $\tau=10^{-\mathrm{SNR}_\mathrm{max} / 10}$ that determines the maximum allowed SNR \cite{wang2019alternating}:
\begin{equation}
    \mathcal{L}_\mathrm{SNR}({y}, \hat{y})
    =     10\log_{10}\left(
    {\|{y}-\hat{y}\|^2 + \tau \|{y}\|^2}
    \right)
    \label{eq:snr}
\end{equation}
and the loss $\mathcal{L}_0$ for ``inactive'' separated sources is
\begin{equation}
    \mathcal{L}_0({x}, \hat{y})
    =10\log_{10}
    \left(
    {\|\hat{y}\|^2 + \tau \|{x}\|^2}
    \right).
    \label{eq:snr0}
\end{equation}
Note that equation (\ref{eq:snr0}) is equivalent to $\mathcal{L}_\mathrm{SNR}$ (\ref{eq:snr}), with the reference ${y}$ set to 0 and the thresholding performed using the mixture ${x}$ instead of the reference source ${y}$.

Thus, separated sources are permuted to match up with either active reference sources ${\bf s}_a$ or all-zeros ``references''. If a separated source is paired with an all-zeros reference, the $\mathcal{L}_0$ loss function (\ref{eq:snr0}) is used, which minimizes the source's power until the ratio of mixture power to separated source power drops below a threshold. When the inactive loss $\mathcal{L}_0$ is small, this threshold prevents the large inactive loss gradients from dominating the overall loss. We found that $30$ dB was a good value for $\mathrm{SNR}_\mathrm{max}$ in order to set $\tau$.

\section{Evaluation Metrics}

\begin{table*}[t]
    \centering
    \caption{Separation results for baseline FUSS model in terms of single-source SI-SNR (1S) and multi-source SI-SNRi (MSi) for various source counts, and source-counting accuracy for under, equal, and over-separation.}
\begin{tabular}{ccccccccccc}
& & & \multicolumn{1}{c}{\bf 1S} & \multicolumn{4}{c}{\bf MSi vs source count} & \multicolumn{3}{c}{\bf Source count rate} \\
\cmidrule(lr){4-4} \cmidrule(lr){5-8} \cmidrule(lr){9-11}
{\bf Eval} & {\bf Split} & {\bf Train} & {\bf 1} & {\bf 2} & {\bf 3} & {\bf 4} & {\bf 2-4} & {\bf Under} & {\bf Equal} & {\bf Over} \\
\midrule
\multirow{ 4 }{3pt}{Dry}
& \multirow{ 2 }{3pt}{Val}
& Dry  & 34.2 & 9.9 & 11.1 & 8.8 & 9.8 & 0.23 & 0.61 & 0.16 \\
& & Rev  & 34.4 & 9.4 & 8.8 & 8.2 & 8.7 & 0.32 & 0.51 & 0.17 \\
\cmidrule(lr){2-11}
& \multirow{ 2 }{3pt}{Test}
& Dry  & 35.5 & 11.2 & 11.6 & 7.4 & 9.8 & 0.25 & 0.60 & 0.15 \\
& & Rev  & 38.4 & 10.9 & 9.0 & 7.7 & 9.0 & 0.32 & 0.54 & 0.15 \\
\midrule
\multirow{ 4 }{3pt}{Rev}
& \multirow{ 2 }{3pt}{Val}
& Dry  & 35.1 & 10.2 & 11.8 & 10.2 & 10.7 & 0.30 & 0.59 & 0.12 \\
& & Rev  & 58.4 & 11.5 & 12.1 & 11.9 & 11.9 & 0.39 & 0.51 & 0.11 \\
\cmidrule(lr){2-11}
& \multirow{ 2 }{3pt}{Test}
& Dry  & 36.9 & 10.7 & 11.9 & 7.7 & 9.9 & 0.29 & 0.60 & 0.12 \\
& & Rev  & 65.8 & 12.7 & 11.6 & 10.2 & 11.4 & 0.35 & 0.57 & 0.08 \\
\bottomrule
\end{tabular}
    \label{tab:sep}
    \vspace{-10pt}
\end{table*}

To evaluate performance, we use scale-invariant signal-to-noise ratio (SI-SNR) \cite{LeRoux2018a}. SI-SNR measures the fidelity between a target ${y}$ and an estimate $\hat{y}$ within an arbitrary scale factor in units of decibels:
\begin{align}
    \text{SI-SNR}({y}, \hat{y})
    & =
    10 \log_{10}
    \frac
    {\| \alpha {y} \|^2}
    {\| \alpha {y} - \hat{y}  \|^2}
    \nonumber
    \\ & \approx 
    10 \log_{10}
    \frac
    {\| \alpha {y} \|^2 + \epsilon}
    {\| \alpha {y} - \hat{y} \|^2 + \epsilon}
    \label{eq:sisnr_a}
\end{align}
with $\alpha = \textrm{argmin}_{a} \| a {y} - \hat{y}\|^2 =  \frac{{y}^T \hat{y}}{\|{y} \|^2} \approx \frac{{y}^T \hat{y}}{\|{y}  \|^2 + \epsilon }$, where a small positive stabilizer, $\epsilon$, is used to avoid singularity. 
We found that this implementation of SI-SNR can lead to inaccuracies in the context of FUSS, especially in the case of {\it under-separation}, where there are fewer non-zero source estimates than there are references. In this case
it produces optimistic scores when $\hat{y}$ is close to 0 due to imprecision in the estimated scale $\alpha$. Our initial reported results on FUSS \cite{turpault:hal-02891700} used this overly optimistic measure. To correct for this, we use an alternate formulation,
\begin{align}
    \text{SI-SNR}({y}, \hat{y})
    & =
    10 \log_{10}
    \frac{{\rho}^2({y}, \hat{y}) }{1 - {\rho}^2({y}, \hat{y}) }  
    \nonumber
    \\& \approx
    10 \log_{10}
    \frac{{\rho}^2({y}, \hat{y}) + \epsilon }{1 - {\rho}^2({y}, \hat{y}) + \epsilon},
    \label{eq:sisnr_p}
\end{align}
where ${\rho}({y}, \hat{y}) = \frac{{y}^T \hat{y}}{\|{y}\|\|\hat{y}\| } \approx \frac{{y}^T \hat{y}}{\|{y}\|\|\hat{y}\| + \epsilon}$ is the cosine similarity between ${y}$ and $\hat{y}$, with stabilizer $\epsilon$.
This is equivalent to (\ref{eq:sisnr_a}) when $\epsilon = 0$, but is more accurate when $\epsilon > 0$.  In our experiments we use $\epsilon=10^{-8}$.

To account for inactive (i.e.\ zero or near-zero) references and estimates, we use the following procedure during evaluation. Separated sources are aligned to reference signals by maximizing SI-SNR with respect to a permutation matrix between estimates and references. Then, estimate-reference pairs are discarded that either have an all-zeros reference, or an estimate with power that is 20 dB below the power of the quietest non-zero reference source. To measure how prevalent this is, we compute the proportion of the examples in the following three categories, which are the three right-most columns of table \ref{tab:sep}:
\begin{enumerate}
    \item Under-separated: fewer non-zero estimates than non-zero references.
    \item Equally-separated: equal number of non-zero estimates and references.
    \item Over-separated: more non-zero estimates than non-zero references.
\end{enumerate}
Note that due to the mixture consistency layer (\ref{eq:mixcon}), the error caused by under-separation is still accounted for, since each under-separated estimate will contain some content of at least one other active non-zero source. From the confusion matrices (Figure \ref{fig:confusion}), the under, equal, and over-separation rates are given by the normalized sums of the lower triangle, diagonal, and upper triangle, respectively.

\section{Experimental Results}



\begin{figure}[t]
    \centering
    \includegraphics[width=0.49\linewidth]{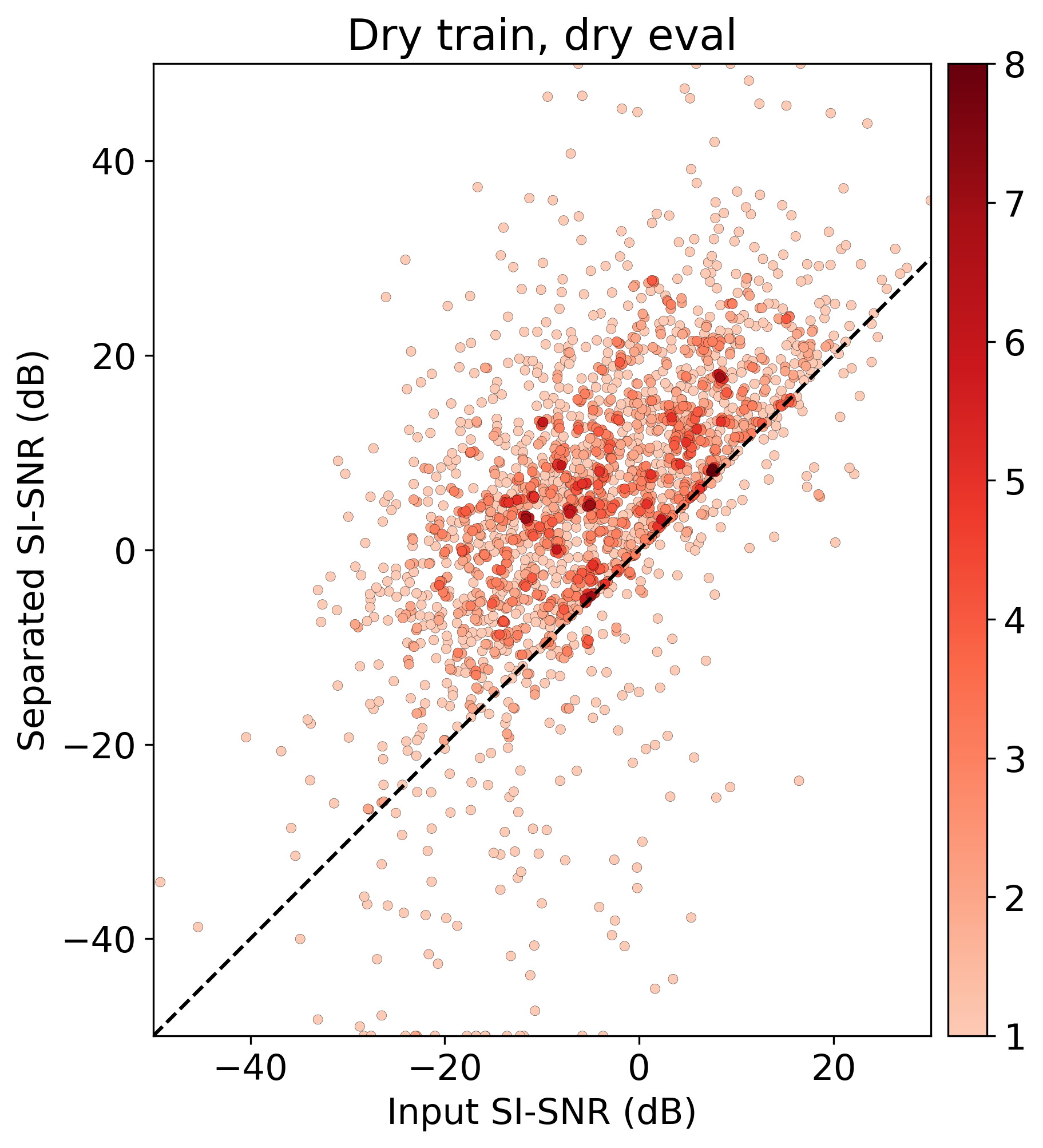}
    \includegraphics[width=0.49\linewidth]{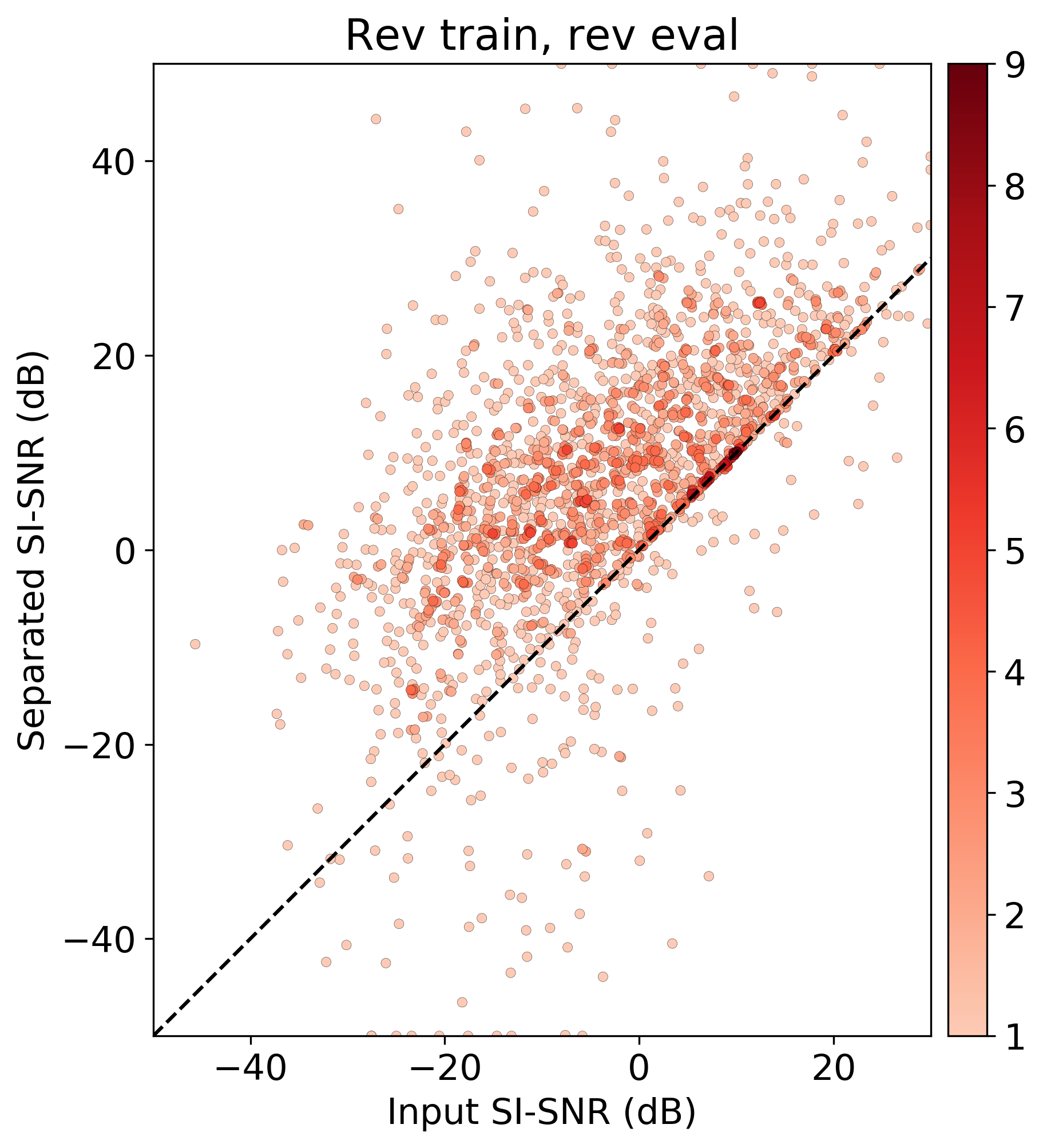}
    \caption{Scatter plots of model performance for examples with 2-4 sources.}
    \label{fig:scatter}
    \vspace{-10pt}
\end{figure}

Table \ref{tab:sep} shows the evaluation results in terms of single-source SI-SNR (1S), multi-source SI-SNRi (MSi), and source counting rates for the baseline separation model on both dry and reverberant validation and test sets of FUSS. Note that the model achieves higher scores on the reverberant sets. This is somewhat unexpected, since other separation models in specific domains, such as speech, often perform worse in reverberant conditions [TODO: citation]. We hypothesize that speech is easier to separate based on its reliable spectral characteristics, whereas for arbitrary sounds, with less predictable spectral patterns,  models may learn to depend on other cues for separation, such as the differing reverberation pattern for each source.
Figure \ref{fig:scatter} shows scatter plots for input SI-SNR versus separated SI-SNR on dry and reverberant test sets for matched training, which visualize the expected improvement for matched models on both dry and reverberant test sets.

\begin{figure}[t]
    \centering
    \includegraphics[width=0.49\linewidth]{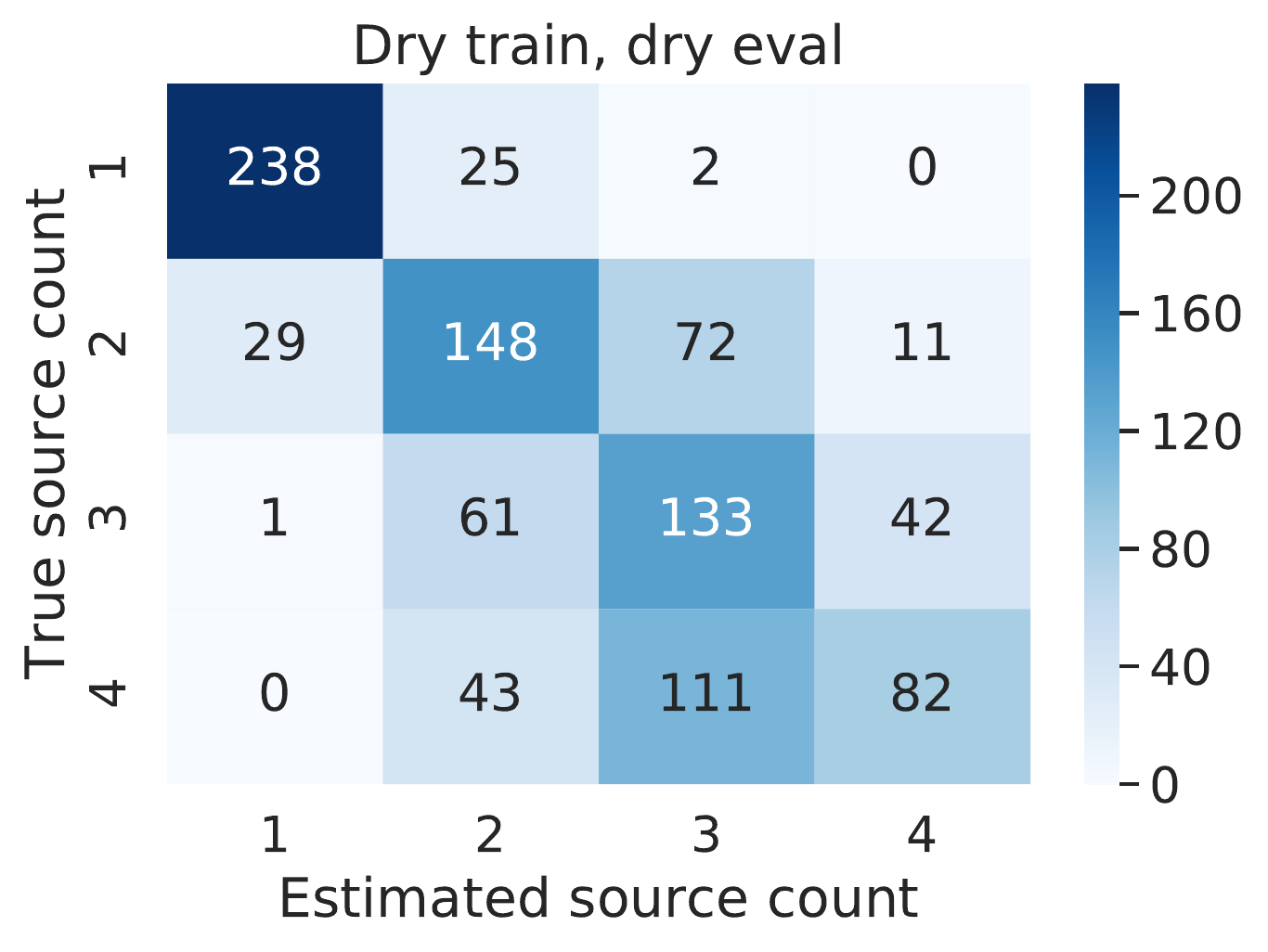}
    \includegraphics[width=0.49\linewidth]{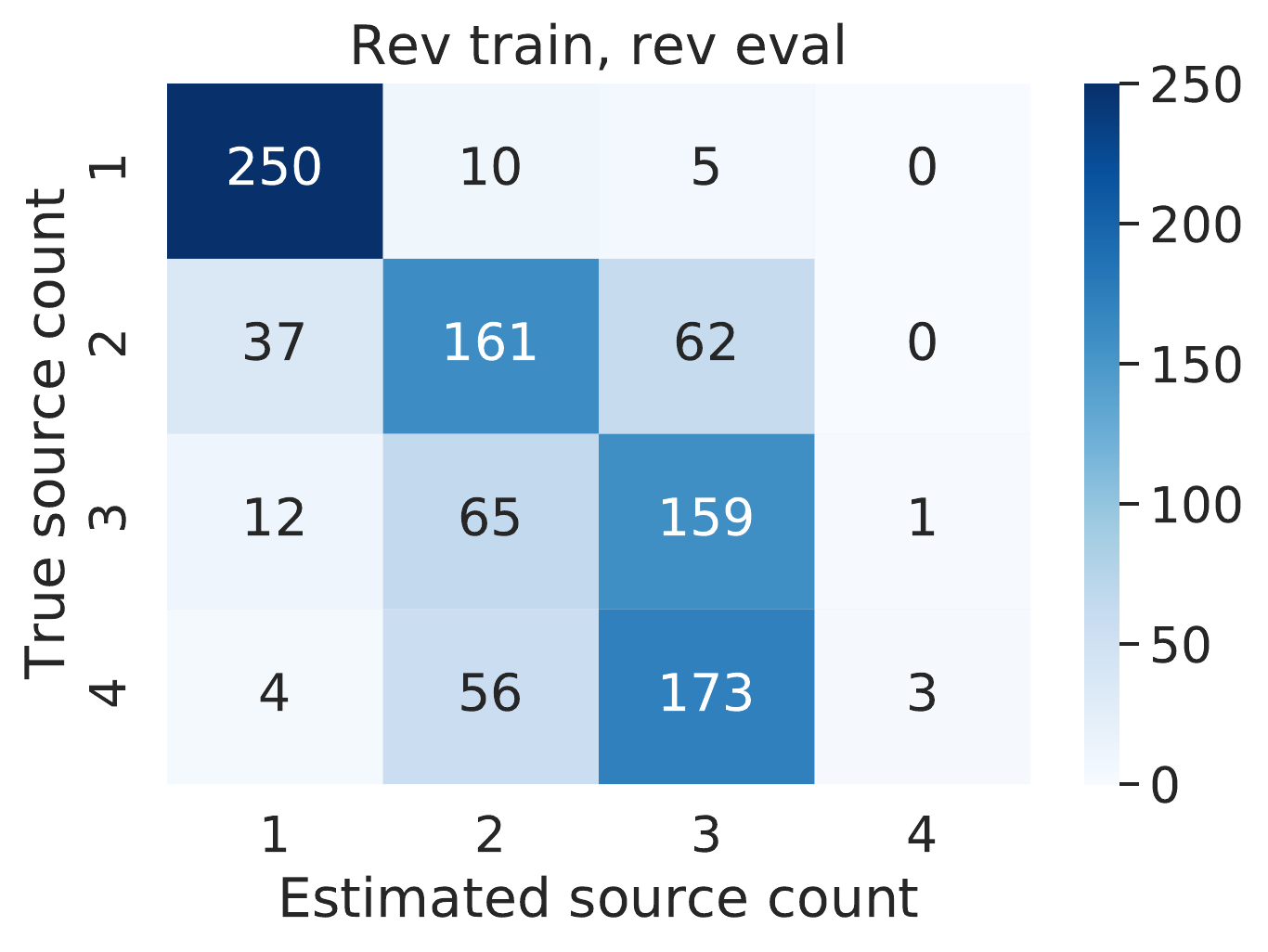}
    \caption{Confusion matrices for source counting, without reverberation (left) and with reverberation (right).}
    \label{fig:confusion}
    \vspace{-10pt}
\end{figure}

As indicated by the source counting proportions, the baseline model exhibits both under-separation and over-separation, tending to under-separate more often than over-separate, which can be seen in the confusion matrices shown in Figure \ref{fig:confusion}. Incorporating mechanisms or regularizations that help avoid these issues is an interesting avenue of future work. One initial finding in another work \cite{wisdom2020unsupervised} is that increasing the number of output sources of the model from 4 to 8, and using mixtures-invariant training (MixIT) \cite{wisdom2020unsupervised} helps to avoid under-separation. However, those models seem to suffer more from over-separation, often struggling to reproduce single-source inputs (i.e., achieve low 1S scores), especially if no supervised training examples of single sources are provided.

\section{Conclusion}
We have presented an open-source dataset for universal sound separation with variable numbers of sources, along with a baseline model that achieves surprisingly good performance at this difficult task.
Future work will include exploring other mechanisms to avoid under- and over-separation. Incorporating class labels from FSD50K is another interesting avenue of further research. We also plan to release the code for our room simulator and extending its capabilities to address more extensive acoustic properties of rooms, materials with different reflective properties, novel room shapes, and so on.

\bibliographystyle{IEEEbib}
\bibliography{refs}

\end{document}